# Mn Solid Solutions in Self-Assembled Ge/Si (001) Quantum Dot Heterostructures

J. Kassim[1], C. Nolph[1], M. Jamet[2], P. Reinke[1], J. Floro[1]

[1]Department of Materials Science and Engineering, University of Virginia, Charlottesville, 22904, USA

[2]Institut Nanosciences et Cryogénie/SP2M, CEA-UJF, F-38054 Grenoble, France

Heteroepitaxial $Ge_{0.98}Mn_{0.02}$ quantum dots on Si (001) were grown by molecular beam epitaxy. The standard Ge wetting layer-hut-dome-superdome sequence was observed, with no indicators of second phase formation in the surface morphology. We show that Mn forms a dilute solid solution in the Ge quantum dot layer, and a significant fraction of the Mn partitions into a sparse array of buried, Mn-enriched silicide precipitates directly underneath a fraction of the Ge superdomes. The magnetic response from the ultra-thin film indicates the absence of robust room temperature ferromagnetism, perhaps due to anomalous intermixing of Si into the Ge quantum dots.

In the last decade, spintronics research has experienced rapid growth, and many of the essential elements of spintronics devices are now in place [1–4]. However, while great progress has been made in some areas, other areas still suffer from the lack of materials "building blocks". Dilute ferromagnetic semiconductors [1], [3], which are highly desirable due to their ability to interface with charge-driven semiconductor technology, remain challenging. This is especially true for group IV semiconductors, due to their compatibility with Si-technology, where epitaxial processes offer potential advantages in lowered interface scattering, reduced current and spin polarization losses, and longer spin lifetimes [5].

While many studies of Mn-doped epitaxial Ge films on Ge(001) have been performed, [6–9], much less has been done on magnetic doping of Group IV quantum dots (QDs) [10–12]. Heteroepitaxial QDs can spatially localize magnetic moments [13], and possibly increase $T_c$ due to hole confinement [14]. Mn incorporation in strain-induced Ge quantum dots poses a particular challenge, since formation of highly-metastable Mn solutions requires low growth temperatures, whereas quantum dot self-assembly invariably requires elevated growth temperatures. Despite these divergent requirements on Mn mobility, Xiu, et al., recently showed remarkably robust ferromagnetism (FM), localized to Mn-doped Ge QDs on Si (001), which could be controlled with an electric field applied to a gate [10]. In recent work [15] we attempted to reproduce this work and provide better understanding of the Ge:Mn self-assembly process. The results we obtained were rather different, both in terms of magnetic response and in the surface-morphological and phase evolution.

To analyze the disparity in results, and to rationalize the magnetic response, it is therefore critical to assess how Mn chemically partitions in the Ge/Si (001) QD system. Here we provide detailed cross-section transmission electron microscopy (TEM), and energy dispersive analysis of x-rays (EDX) measurements to characterize the structure and composition of heteroepitaxial Ge/Si (001) QDs with 2 at% nominal Mn composition. This sample, chosen since it exhibits no indications of second phase formation in atomic force microscopy (AFM), was one of a series of samples where nominal Mn composition was varied from 0–10 at% to examine the effect of Mn on surface morphology and magnetism [15]. The film was grown by molecular beam epitaxy (MBE) using magnetron sputter deposition of Ge in 4.3 mTorr of getter-purified Ar with thermal co-evaporation of Mn from a BN crucible in an effusion cell. The MBE base pressure was $1 \times 10^{-10}$ Torr. Undoped Si (001) substrates were chemically cleaned and passivated with a sub-oxide that is desorbed in situ at 800ºC prior to 50 nm Si buffer growth that results in a smooth, 2x1 reconstructed surface; details are described elsewhere [15]. A 7.3 ML thick Ge:Mn film was grown at 450ºC and 0.1 ML/min. The film was analyzed by SIMS using $Cs^+$ ions at 5 keV and 60° off-normal incidence. Surface morphology was characterized *ex situ* using tapping-mode AFM. TEM was employed for high-resolution imaging and compositional analysis using EDX. Magnetic properties were measured using a superconducting quantum interference device (SQUID) magnetometer with the external field applied normal to the film. The samples were carefully handled to prevent magnetic contamination. The portion of the film for SIMS analysis was first capped with 100 nm amorphous Si at room temperature after previous air exposure of the sample. Depth profiles for Ge and Mn were obtained by sputtering entirely through the films and into the Si buffer. The Mn signal was calibrated versus a Mn+ implanted Si wafer standard. The total Ge integrated over depth was $4.96 \times 10^{15}$ at/cm$^2$ while Mn was $1.05 \times 10^{14}$ at/cm$^2$, for a nominal Ge:Mn composition of 2.1 at% Mn.

Even with a 2% Mn flux during Ge growth, we still observe self-assembly of the quantum dot morphologies ubiquitous to the Ge/Si (001) system [12, 13, 14] as shown in FIG. 1: "hut cluster" islands ("H"), dome clusters ("D") and larger superdomes ("SD") [16]. No atypical morphologies are observed that would indicate second phase formation. By comparison, second phase formation does become obvious in AFM for $Ge_{0.95}Ge_{0.05}$ films, where small protrusions emerge from Ge SD and D dots [15] and for $Ge_{0.9}Ge_{0.1}$ films, where second phase precipitates dominate the overall surface morphology [to be published]. Hence it is of great interest here to understand where the Mn is located, in the absence of obvious surface precipitates. Our analysis of the SIMS depth profiles demonstrated that $7 \times 10^{13}$ Mn atoms reside within the Ge, implying that *the average composition of the Ge layer is about 1.5 at% Mn* (this assumes no Si alloying, but see below). The remaining $\leq 3 \times 10^{13}$ Mn is in the Si substrate, either due to diffusion during growth, or due to ion mixing during SIMS analysis. We show below that Mn diffusion into Si indeed occurs during growth of the Ge:Mn QD layer.

FIG. 2 shows cross-section TEM micrographs of a Ge SD. Internal clustering of Mn within the relatively large, diamond cubic Ge QD is not observed in this island, or in any of the 30 D/SD dots we examined in this sample. EDX using a 2 nm probe assesses the composition of the superdome at point 1 in FIG. 2(a) as $Mn_{0.05}(Ge_{0.54}Si_{0.41})$. In this SD, and in other SD islands, EDX consistently measures local Mn contents of order 1-5 at%, somewhat enriched compared to the SIMS area-averaged concentration, perhaps indicating preferential incorporation in superdomes. In addition, EDX suggests significant intermixing of Ge and Si in the SD island, which was unexpected at these temperatures. FIG. 2(a) also shows the presence of a buried structure beneath the Ge SD, not detectable by AFM, where the cross-section typically presents as trapezoidal. Of 20 SDs surveyed, 4 exhibited similar buried structures; none were observed under Ge domes. EDX at point 2 shows there is at least 3x enrichment of Mn in the buried region compared to the SD, i.e., composition of at least 15 at% Mn in Si. There is no evidence from high-resolution imaging or diffraction that these structures are anything other than diamond cubic.

FIG. 3 shows TEM micrographs exhibiting Ge D and SD islands, where the latter exhibits a different type of buried structure. EDX using a 2 nm probe assesses the composition of the superdome at point 1 in FIG. 3(b) as $Mn_{0.01}(Ge_{0.45}Si_{0.54})$. The buried structure, and two similar structures (not shown) display sharp interfaces and highly regular Moiré fringes with spacing of order 1 nm, and are typically more equiaxed than the structure type shown in FIG. 2. The Moiré indicates that these regions are not isostructural to Si, i.e., they are Mn silicide phases [17]. However, their crystal structure has not been indexed due to a paucity of diffraction spots (suggesting we are not along a precipitate zone axis). Based on the Moiré patterns observed from several such structures, they exhibit a range of lattice orientations and/or crystal structures. EDX at point (2) in FIG. 3(b) confirms that the buried structure is 20-30x more enriched in Mn than the Ge QD, but the surrounding Si matrix precludes better quantification. Other films we have grown with larger Mn content exhibit both $MnSi_{1.75-x}$ and $MnSi$ phases [to be published].

FIG. 3(d) shows a high angle annular dark field (HAADF) image showing a precipitate that appears to be growing up into the Ge QD.

We can estimate the number of Mn atoms in the buried precipitate in FIG. 3(c) by approximating the shape as a semi-ellipsoid of revolution and assuming the phase is $MnSi_{1.75}$. This yields $1.6 \times 10^5$ Mn atoms, which, given the measured Mn in the Si from SIMS, implies that Mn has been scavenged from a surface capture area of about $5 \times 10^5$ $nm^2$. From AFM measurements of QD areal density, there are about 7 superdomes in this capture area. The propensity to take Mn from such a large area leads us to question why the Ge SD's directly above the precipitates still retain significant Mn. We speculate that Mn incorporated in the Si in the early stages of growth, e.g., during formation of the Ge wetting layer, where intermixing might be more facile. This period occupies the first third to half of the film growth. Once quantum dots form, diffusing surface Mn is then captured in the dots, especially the large superdomes, which appear to preferentially nucleate atop the precipitates.

Interestingly, there is no indication of nm-scale Mn clustering *within* QDs, despite the well-known tendency for Mn to undergo a spinodal-like decomposition to form enriched 3-5 nm clusters in Ge at much lower growth temperatures [18–20]. It seems likely that the free energy curve for the diamond cubic solution phase exhibits a shallow spinodal at temperatures < 200ºC, so that at 450ºC, mixing entropy overcomes the positive heat of mixing, suppressing the spinodal. Paradoxically, growth at high temperatures can actually lead to more homogeneous incorporation of Mn in Ge, at least in ultra-thin films.

The large Mn-enriched regions showing diamond cubic crystal structure (see FIG. 2), have interfaces that are much sharper than a simple diffusion profile, and exhibit a tendency towards faceting. Given that the Mn content is lower than any equilibrium Mn-Si phase, this suggests the formation of a metastable phase, perhaps involving Mn in interstitial sites, such as had been suggested in the literature for the Mn-Ge system [21]. When some critical concentration is attained, nucleation of a crystallographically-distinct silicide phase then occurs. Our experiments with similar Ge films having higher Mn contents than used here, reported elsewhere, show that higher supersaturation drives copious silicide nucleation, followed by rapid precipitate growth both above and below the Si substrate, consistent with the nascent emerging precipitate in FIG. 3(d). Once precipitates emerge from their host quantum dots onto the exposed growth surface, they grow rapidly, scavenging all the available Mn due to their lower chemical potential vis-à-vis the metastable solutions.

The magnetic behavior is shown in FIG. 4. Unfortunately, definitive interpretation of this data in terms of the magnetism associated with the Mn is not possible. The maximum possible magnetization from the Mn is about $1 \times 10^{-6}$ emu [footnote: This assumes complete alignment of 5 $\mu_B$/Mn moments and a 4x6 mm sample size). While this small signal is well within SQUID sensitivity ($1 \times 10^{-8}$ emu), we are limited by the need to subtract out signals from the bulk substrate, including both Si diamagnetism and any (para)magnetic impurities associated with the substrate. We subtracted the carefully mass-normalized data from the 0% Mn control sample in order to account for any ferromagnetic impurities in the Si. However, since there is less than 0.2 ML of Mn (relative to the Si (001) planar density), the substrate contribution will constitute more than 99% of the raw signal. Hence the subtraction is extremely sensitive, and we obtain an M-H loop at 5K, shown in FIG. 4(a), in which $M_s$ is much too large to attribute only to Mn, suggesting the presence of some additional paramagnetic impurity content that differs from sample to sample. While additional *ad hoc* subtractions can be employed, the results are meaningless. At higher temperatures, e.g., from above 40 K, we do see "S-shaped" M-H curves persisting to room temperature that could result from the Mn, with saturation moments of about 1 $\mu_B$/Mn, but the noise, which has its origins in the subtraction process, makes interpretation questionable. FIG. 4(b) shows the zero field cooled (ZFC) and field cooled (FC) curves using 200 Oe for the 2% sample, and for the 0% Mn control sample, with no subtractions. The ZFC-FC curves for the control sample overlay completely. For the 2% sample, there is consistently a small separation of the FC curve above the ZFC from 5 K to at least 140 K of about $1 \times 10^{-8}$ emu, right at the noise threshold of the SQUID. This could indicate a small ferromagnetic component attributable to the Mn, but would only represent about 1% of Mn atoms contributing. Given the very small signal, it is only possible to say that *robust room-temperature ferromagnetism arising from the majority of the Mn atoms in not observed*.

Clearly, the magnetic response found here is quite different from the results of Xiu et al. [10]. Nominally, our growth conditions are quite similar to theirs, and in other experiments reported elsewhere, we bracket the range of likely Mn contents [15]. Furthermore, from a structural standpoint, our 2% Mn film appears quite similar to that shown by Xiu, et al.; both groups observed Mn solid solutions in Ge QDs and buried in diamond cubic Si regions immediately below the QDs. To examine the role of impurities, we used SIMS analysis of Ge and Si calibration samples grown under clean conditions in our MBE, which do not detect transition metal impurities Ni, Co, and Fe down to the sensitivity of SIMS, less than 1 ppm. Carbon and oxygen are observed, at levels less than 50 ppm. These levels, which are 3 orders of magnitude smaller than the Mn content, might delicately affect carrier transport, but it seems unlikely that complete suppression of ferromagnetism can be ascribed to these impurities.

The Ge:Mn QD solution phase grown here may be more dilute than in Xiu, et al.'s work, reducing hole density and increasing coupling distances. However, increasing the Mn concentration in the MBE growth flux leads to copious second phase formation easily detected by AFM, so we effectively are bounded by this Mn content. EDX local probe measurements indicated that the Ge quantum dots are intermixed with Si to about 50 at%. Comparisons of magnetic behavior in amorphous Ge:Mn and Si:Mn films shows that Mn tends to occupy interstitial positions in the latter and substitutional positions in the former. High coordination interstitial occupation is suggested to enhance *p-d* hybridization that quenches the local Mn moment [22], [23]. The apparent Ge-Si alloying is very surprising, given the low growth temperature (alloying is usually observed only above 550-600 ºC), and since the small QD size we observe is more consistent with pure Ge dots (deliberate growth of alloy dots at 50-50 composition yield structures two orders of magnitude larger in volume). Nonetheless, we cannot confirm that the mixing results from an artifact of the measurement or the TEM specimen preparation.

In conclusion, we have synthesized epitaxial self-assembled heteroepitaxial QD's by MBE co-deposition of nominal $Ge_{0.98}Mn_{0.02}$ films. The 2 at.% Mn has no perceptible effect on the QD's morphology. A Mn solution phase forms in the Ge QDs, especially within the superdomes, without any indication of nanoscale phase separation. About 30% of the total Mn partitions into a sparse array of buried precipitates in the Si substrate, forming under about one third of the superdomes, that are highly enriched in Mn, resulting in both a metastable diamond cubic phase, and crystallographically distinct silicide phases at even higher levels of Mn enrichment. The extremely small magnetic signal from the tiny amount of Mn, superimposes on a much larger substrate signal, making interpretation of the magnetic phases very difficult. This would be true even if all the Mn atoms were ferromagnetically coupled. There is some indication in ZFC-FC data for a ferromagnetic coupling associated with Mn, but there is no clear evidence of robust room temperature ferromagnetism, perhaps due to the presence of Mn-Si bonds in both the substrate and the QDs.

We gratefully acknowledge Dr. Alline Myers for the TEM work performed in part at the NIST Center for Nanoscale Science and Technology, Dr. Joshua Schumacher (CNST) and Trevan Landin (FEI Company) for the FIB TEM sample preparation, Dr. Stephen Smith (Evans Analytical Group) for his help with SIMS characterization and the help of Richard White on TEM. This work is supported by the National Science Foundation under grant number DMR-0907234.

**Figure Captions**

FIG. 1. Surface morphology from AFM showing (a) phase image to highlight faceting and (b) true topography.

FIG. 2. XTEM micrograph along the [110] zone showing a Ge superdome. Both the SD and the buried region beneath contain Mn, according to EDX scans in (b) and (c).

FIG. 3. XTEM micrographs along the [110] zone, showing (a) a Ge dome, (b) a Ge superdome with a buried precipitate that is enlarged in (c) to highlight the Moire pattern. Panel (d) shows a HAADF image from a different area of the TEM sample, where a buried precipitate is growing up into the Ge quantum dot.

FIG. 4. (a) M-H loops measured by SQUID at various temperatures, compared with the largest possible signal (solid red line) if every Mn atom is ferromagnetically coupled, using a simple Langevin behavior. (b) ZFC-FC data for the control sample and 2% specimen.

Figure 1

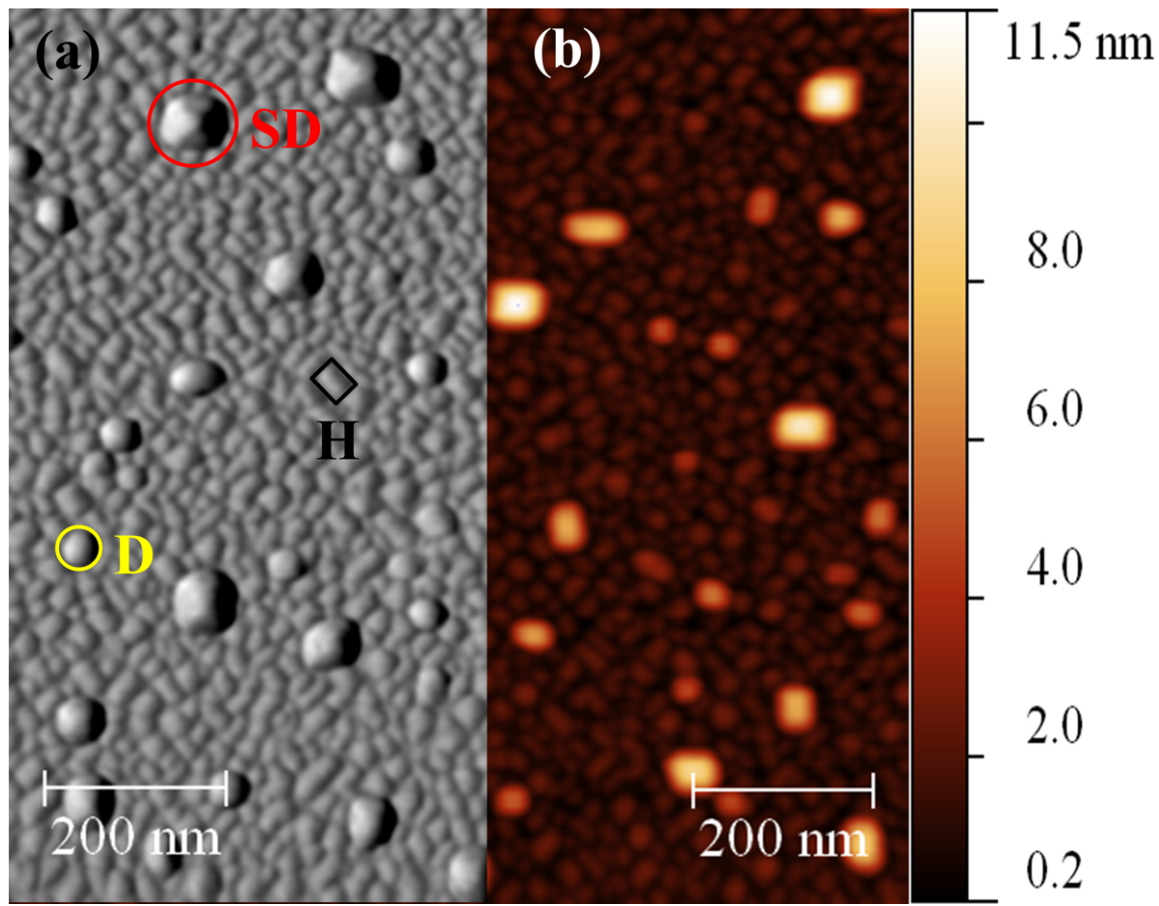

Figure 2

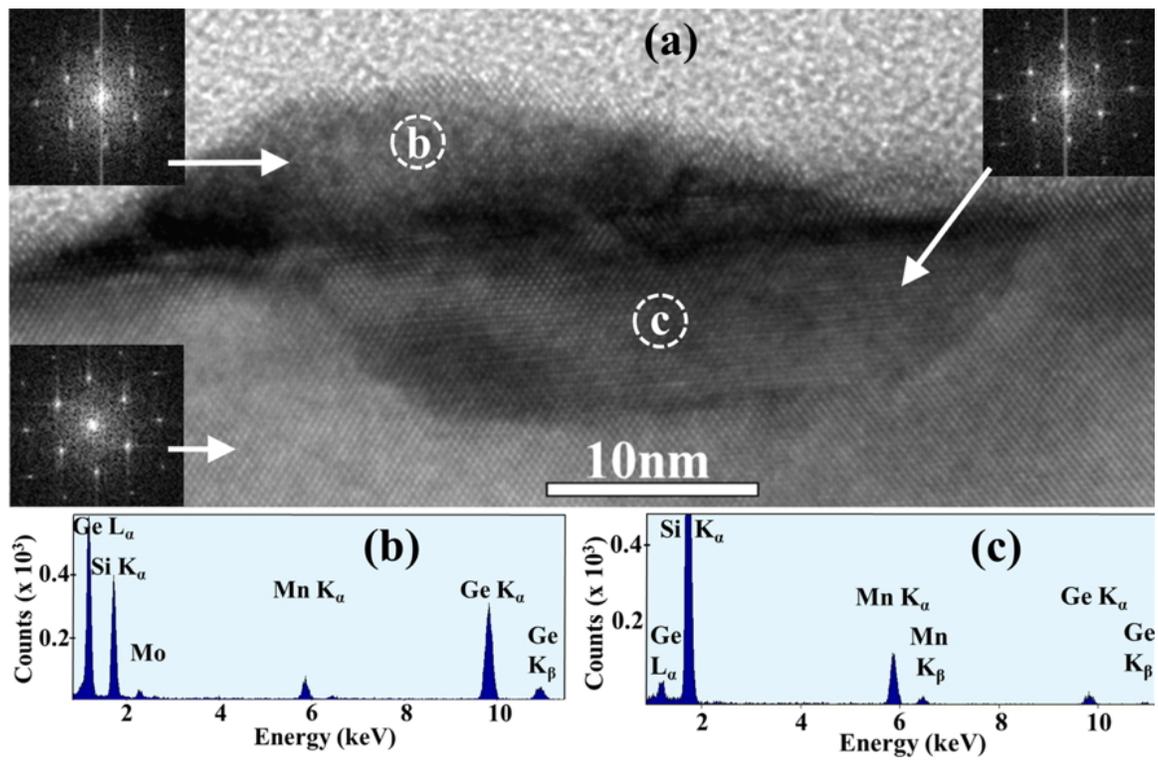

Figure 3

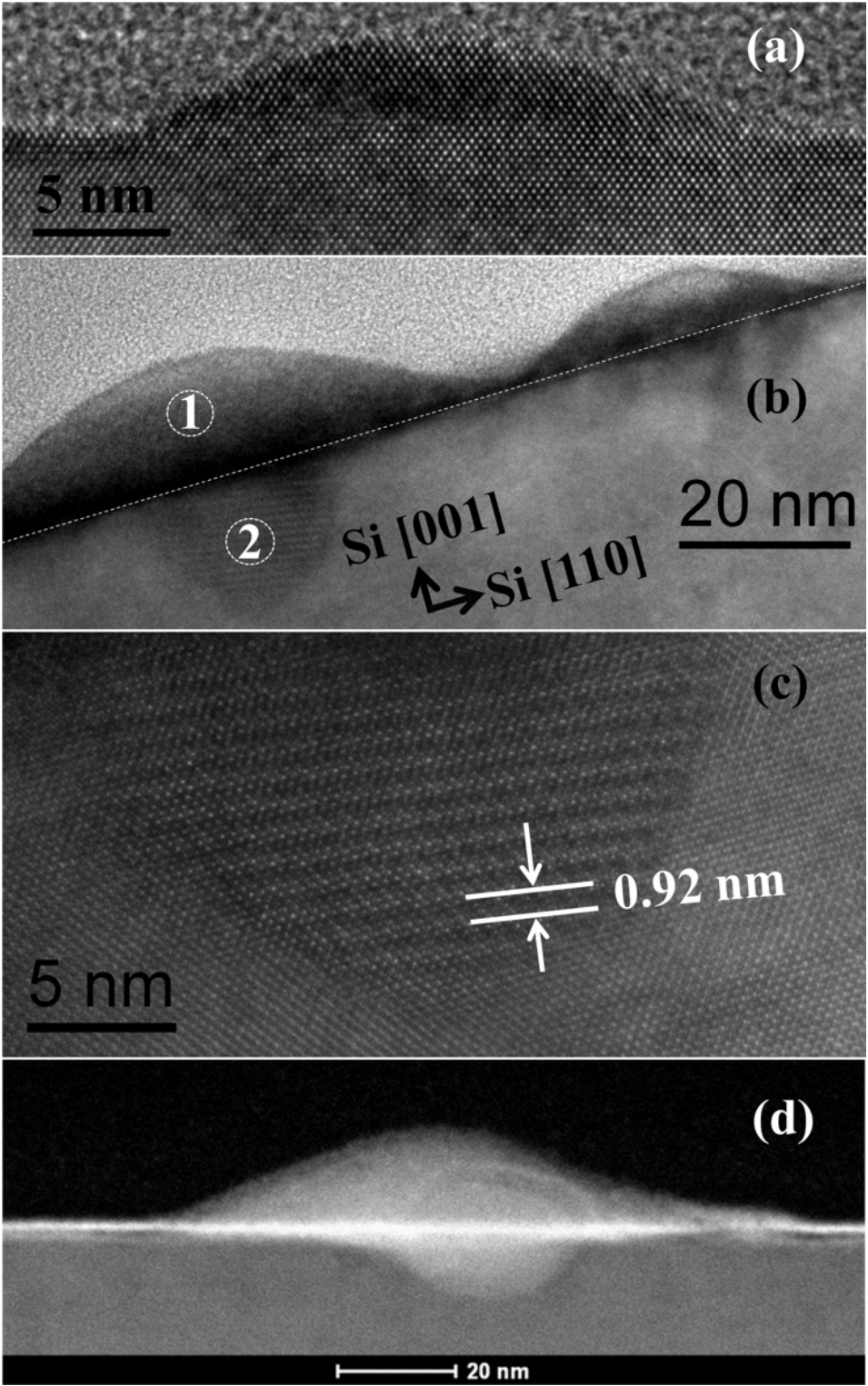

Figure 4

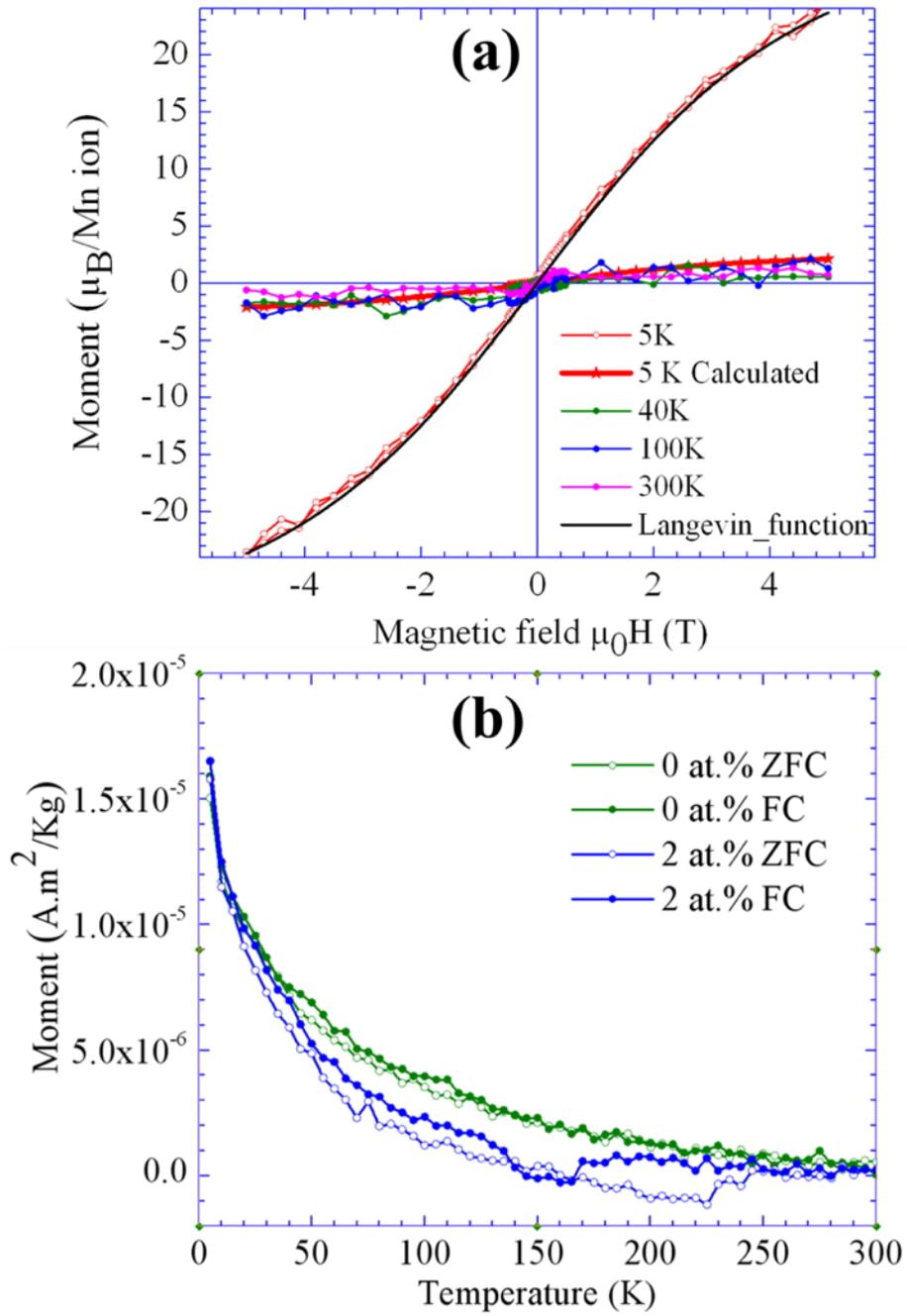